\documentclass[runningheads]{llncs}
\usepackage{graphicx}
\usepackage{xcolor}
\usepackage{hyperref}

\newcommand{\app}[1]{midiVERTO{#1}}

\begin{document}
\title{
\app{}: A Web Application to Visualize Tonality in Real Time%
\thanks{This research project has received funding from the UNIL-EPFL dhCenter, an interdisciplinary research platform set up between Universit\'e de Lausanne (UNIL) and \'Ecole Polytechnique F\'ed\'erale de Lausanne (EPFL), Switzerland.}
}


\author{
  Daniel Harasim \inst{1} \and
  Giovanni Affatato \inst{2} \and
  Fabian C. Moss \inst{1,3}
}

\authorrunning{D. Harasim et al.}
%
\institute{
  \'Ecole Polytechnique F\'ed\'erale de Lausanne, Lausanne, Switzerland\\ 
  \email{daniel.harasim@epfl.ch}
  \and 
  Politecnico di Milano, Milan, Italy\\ \email{giovanni.affatato@mail.polimi.it}
  \and 
  University of Amsterdam, Amsterdam, The Netherlands\\ \email{fabian.moss@uva.nl}
}


\maketitle              
\begin{abstract}
This paper presents a web application for visualizing the tonality of a piece of music---the organization of its chords and scales---at a high level of abstraction and with coordinated playback.
The application applies the discrete Fourier transform to the pitch-class domain of a user-specified segmentation of a MIDI file and visualizes the Fourier coefficients' trajectories.
Since the coefficients indicate different musical properties, such as harmonic function, triadicity, and diatonicity, the application isolates aspects of a piece's tonality and shows their development in time.
The aim of the application is to bridge a gap between mathematical music theory, musicology, and the general public by making the discrete Fourier transform as applied to the pitch-class domain accessible without requiring advanced mathematical knowledge or programming skills up front.

\keywords{Web application \and Visualization \and Discrete Fourier transform \and Tonality \and MIDI.}
\end{abstract}


\section{Introduction}

Music analysis traditionally requires a high degree of expertise, for example to read and interpret scores in order to derive insights about latent tonal structures in a composition. 
Mathematical music theory provides tools to study musical phenomena with an emphasis on comparing various pieces and styles by generalizing from concrete pieces to more abstract concepts and definitions.
For example, many interesting musical entities can be described as subsets of cyclic groups, such as chords, scales, and repeating rhythms.
In recent years it became increasingly popular among music theorists to study such entities by applying the discrete Fourier transform (DFT) to the pitch domain~\cite{amiotEntropyFourierCoefficients2021,tymoczkoFourierPhasePitchclass2019,yustSchubertHarmonicLanguage2015}, in contrast to the time domain to which the DFT is most commonly applied (e.g., in signal processing and music information retrieval).
For a general description of this method's mathematical details and music-theoretical interpretations see for instance
\cite{amiotMusicFourierSpace2016}.
This usage of the DFT essentially maps subsets of cyclic groups to six complex numbers (i.e., the Fourier coefficients 1 to 6) and many interesting properties, such as evenness \cite{amiotDavidLewinMaximally2007,harasimDistantNeighborsInterscalar2019}, balancedness \cite{milneExploringSpacePerfectly2017}, and diatonicity \cite{yustStylisticInformationPitchclass2019}, can be identified and studied with the Fourier coefficients.

Since the DFT can be applied to music in symbolic formats (e.g., MIDI) without prior interpretation of the musical material by a music theorist, it is well suited for distant-reading approaches in corpus studies, such as the comparison of different pieces' tonal organization at a high level of abstraction \cite{yustStylisticInformationPitchclass2019}.
However, this requires advanced mathematical and computational skills, which often hinders music historians, students, and enthusiasts from engaging with it. Harnessing the power of such complex methods thus remains restricted to only a small group of researchers.

This paper presents the open-source web application \app{} which addresses this problem and aims to bridge the intradisciplinary methodological divide between mathematical music theory, musicology, and the general public.
The application enables its users to visualize the tonality of a piece of music viewed through the lens of the DFT without the requirement to understand its formal details. The app thus contributes to an exchange of knowledge and techniques for music analysis between 
In this spirit, we continue the work by Thomas Noll~\cite{nollInsidersChoiceStudying2019} and Jennifer Harding~\cite{hardingComputerAidedAnalysisTonal2020}
to make DFT analyses of musical entities more easily accessible and reproducible for a broader readership, especially students.

\section{\app{}: features and technical details}

\begin{figure}[t]
    \centering
    \includegraphics[width=0.84\textwidth]{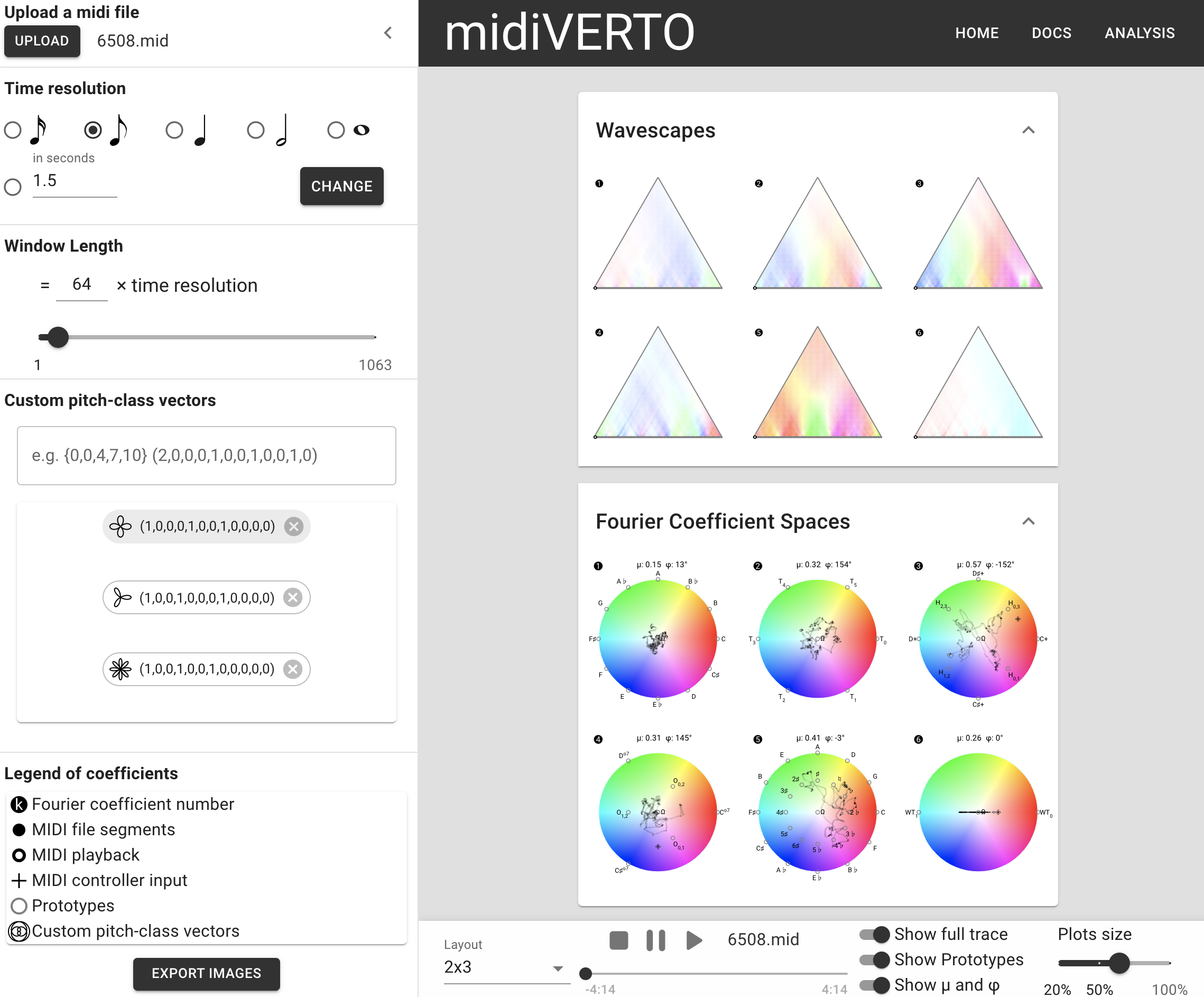}
    \caption{Screenshot of the web application \app{} with three components: input-output on the left, visualizations in the center, and control options at the bottom.}
    \label{fig:screenshot}
\end{figure}

A screenshot of the application is shown in Figure~\ref{fig:screenshot}.%
    \footnote{The app is accessible at
    \url{https://dcmlab.github.io/MIDFT}
    }
Apart from a welcome page (``Home'') and a step-by-step tutorial (``Docs''), the application features an interface consisting of four parts (``Analysis''): an input-output menu on the left, two visualization panels in the center, and a control menu at the bottom.

To visualize the tonal content of a piece encoded in MIDI format, a user can upload the MIDI file and select a duration value that is used to segment the music. For instance, a value of a quarter note leads to a division into many quarter-note segments. 
Durations can be specified either in note values or in seconds.
The application then creates two interlinked visualizations. Note that the user-specified segmentation is only relevant for the visualization in the lower panel. For the one in the upper panel, a fixed segmentation is used to avoid unnecessarily long computation times.

The first visualization shows six \emph{wavescapes}~\cite{viaccozWavescapesVisualHierarchicalinpress}. These are triangular plots closely related to keyscapes~\cite{sappVisualHierarchicalKey2005} and pitch scapes~\cite{lieckModellingHierarchicalKey2020}.
The $k$-th wavescape shows the values of the $k$-th Fourier coefficient in time and at different time scales. 
Magnitudes of the Fourier coefficients are indicated by opacity and thier phases are shown according to a circular color mapping described below, together with the second visualization.
The bottom of a wavescape corresponds to the respective coefficient's values for each segment.
All other horizontal slices take groups of segments into account. The higher the slice, the more segments are considered. The high levels of the $k$-th wavescape thus show the values of the $k$-th Fourier coefficient on large time scales, and the tip of each triangle represents the value of the respective Fourier coefficient for the pitch-class content of the entire piece.

In contrast to the wavescape visualization, which is static, the second visualization is dynamic and coordinated with the playback of the MIDI file.
For each of the six Fourier coefficients, a unit disk in the complex plane is shown and each angle is assigned a color in a circular color mapping.
We call these disks \emph{Fourier coefficient spaces}.
Each disk shows the postions of prototypical pitch-class sets to provide orientation in the space. Those are sets that have high magnitudes in the respective coefficient, such as augmented triads in the third and diminished seventh chords in the fourth coefficient.

Each time point of the uploaded piece is assigned to six complex numbers.
Similar to the wavescape visualization, the coefficient-space visualization can show values according to different time scales, but in contrast to wavescapes only with respect to one time scale at a time.
A time scale corresponds to the length of a sliding window over segments and this length can be set by the user in the input-output menu. For each sliding window, all contained pitch classes are summed into one 12-dimensional pitch-class count vector which is subsequently normalized ($L^1$ norm) and mapped under the Fourier transform. 
The normalization ensures that all coefficients' values are contained inside the unit circle.
Very short window lengths tend to distribute the segments of the piece more evenly throughout the disks. In contrast, long window lengths collapse the segments into a small area, and even into a single point if the window covers the entire piece (i.e,  the topmost point of a wavescape). Therefore, it is beneficial to try different window lengths in order to identify an appropriate level of detail for the concrete piece under analysis, for instance one where the piece moves along smooth paths in some coefficient spaces.

The user can play, pause, and skip through the MIDI file using the control menu at the bottom of the app. While the piece is being played, the coefficient values of the current sliding window are displayed as white dots in the coefficient spaces.
The control menu contains additional options, such as for hiding the prototypical pitch-class sets and for the adjustment of the visualizations' sizes and layouts.
%
A user can also manually input pitch-class multisets or distributions into \app{}. Those can be either typed into a text field in the input-output menu or played through an external MIDI controller, such as a keyboard or music notation software.
This feature can be particularly useful to compare specific chords and scales of interest, for example in educational contexts. 

\app{} was developed as a purely browser-based application in order to minimize maintenance requirements and to facilitate a wide adoption by the community. That is, no installation is required and no server application needs to be maintained since the application is entirely client-based.
This also allows the application to run across platforms and lowers the entry barrier for users that might not be confident with technical installation instructions.
The application is written in javascript using the libraries react for state management, material-ui for adaptive interface design, and Canvas API and SVG markup for visualization.
The source code is publicly available on Github under a GPL3 licence, and the Github Pages service is used to deploy the application.\footnote{Source code available at \url{https://github.com/DCMLab/midiVERTO}}

\begin{figure}[t]
    \centering
    \includegraphics[width=0.6\textwidth]{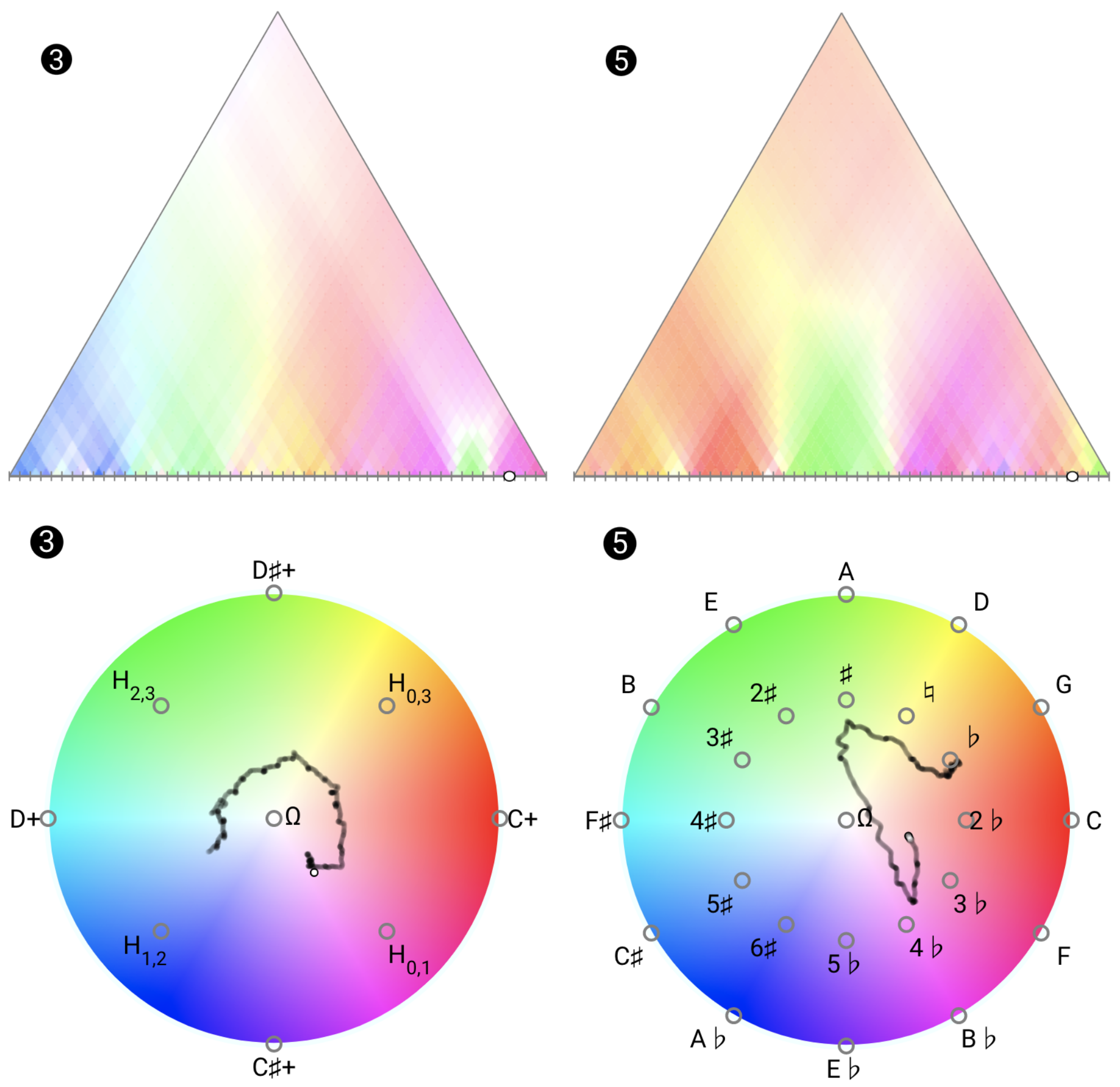}
    \caption{Visualizations of the 3rd and the 5th Fourier coefficient for the \emph{Phantom of the Opera} theme (resolution: 8th note, time window: 300 $\times$ resolution $=37.5$ whole notes). The unit disks are labeled with their prototypes: augmented triads are denoted with a $+$, hexatonic scales by two contained pitch classes (e.g., $H_{2,3}=\{2,3,6,7,10,11\}$), singleton pitch classes by a representative, and diatonic scales by their key signatures.}
    \label{fig:phantom}
\end{figure}

\section{A brief case study}
To showcase \app{}'s capabilities, we use the main theme of the musical \emph{Phantom of the Opera} by A. L. Webber with a resolution of an 8th note and a window length of 300 times the resolution.%
    \footnote{MIDI file taken from 
        \url{https://bitmidi.com/uploads/6508.mid}}
Each point thus represents a length of $37.5$ whole notes.
Central important aspects of the piece's tonality are visualized intuitively in the wavescapes and the Fourier coefficient spaces.
The wavescapes 3 and 5 have the strongest colors among all 6, indicating that diatonic scales as well as (augmented) triads play and important role for the overall harmonic organization of the piece. Therefore, we focus only on these two Fourier coefficients (see Figure~\ref{fig:phantom}).

On the lower levels of the 3rd wavescape, the colors blue -- green -- yellow -- pink ( -- green -- ) pink change in regular time intervals, following the primary key changes of the piece. Tracing this sequence of colors in the third coefficient space, we observe that these colors correspond to the hexatonic scales $H_{1,2}$, $H_{2,3}$, $H_{0,3}$, and $H_{0,1}$, respectively (in clockwise direction; see figure caption for explanation of notation).
The overall harmonic trajectory of the piece thus moves in descending minor thirds, and at the end it briefly moves across the plane to its hexatonic pole (the green interruption in the pink area).

The color progression on the lower levels of the 5th wavescape is roughly orange/red - green - pink, showing that the initial, middle, and final parts of this piece modulate through different keys. Note that the red and pink areas are adjacent to each other but opposite to the green area, again showing a symmetrical organization in terms of keys. Analogous observations can be made by following the path in the 5th coefficient space, because the window size of 300 eighth notes is adjusted to the required abstraction level.

\section{Conclusion}

Interactive applications like \app{} can help to promote mathematical tools for music analysis among a broader audience, such as music theorists, students, and practitioners. They can guide musical intuition and lead to new ways of hearing without the need of studying a complex methodology up front. Furthermore, they can spark interest, generate curiosity, and inspire to learn more about the inner workings of techniques such as the DFT.
%
Moreover, we expect that the application will enable scholars and students of music to employ this powerful method in their work and teaching, and for public engagement. Beyond purely music-theoretical contexts, the application is also suited to teach and study visualization techniques for complex cultural data in the Digital Humanities and Cultural Analytics.

~\\ \noindent \textbf{Acknowledgments.} 
We thank Martin Rohrmeier for his support and guidance as well as Jason Yust and C\'edric Viaccoz for their valuable comments in the development process of the web application.

\bibliographystyle{splncs04}
\bibliography{references}

\end{document}